\magnification=1200
\baselineskip14pt plus1pt minus1pt
\vsize=23truecm
\overfullrule0pt

\font\bff=cmb10
\font\bb=msbm10

\def\R{\hbox{\bb R}}

\def\bs{\bigskip}
\def\ms{\medskip}
\def\ss{\smallskip}
\def\nt{\noindent}

\def\i#1/#2\par{\item{\hbox to .6truecm{#1\hss}}#2}
\def\ii#1/#2/#3\par{\advance\parindent by .8truecm
       \item{\hbox to .6truecm{#1\hss}
         \hbox to 0.6truecm{#2\hss}}#3
          \advance\parindent by -.8truecm}
\def\iii#1/#2/#3/#4\par{\advance\parindent by 1.6truecm
        \item{\hbox to 0.6truecm{#1\hss}
              \hbox to 0.6truecm{#2\hss}
              \hbox to 0.6truecm{#3\hss}}#4
        \advance\parindent by -1.6truecm}


\centerline{\bff A POSITIVE COSMOLOGICAL CONSTANT IN STRING THEORY} 
\ss
\centerline{\bff THROUGH AdS/CFT WORMHOLES}
\ms
\centerline{by}
\ms
\centerline{Brett McInnes}
\ms
\centerline{Department of Mathematics}
\ss
\centerline{National University of Singapore}
\ss
\centerline{10 Kent Ridge Crescent, Singapore 119260}
\ss
\centerline{Republic of Singapore}
\bs\bs\bs

\nt{\bf ABSTRACT:} \ \ There are two important examples of physical systems which violate the strong energy condition : Universes (like, it would seem, our own) with a positive cosmological constant, and wormholes.  We suggest that a positive cosmological constant can be reconciled with string theory by considering wormholes in string backgrounds.  This is argued in two directions : first, we show that brane-worlds with positive cosmological constants give rise to bulk singularities which are best resolved by embedding the brane-world in an AdS/CFT wormhole; and second, for the simplest kind of wormhole in an asymptotically AdS space, we show that the IR stability of the matter needed to keep the wormhole open {\it requires} the presence of a brane-world.  UV stability conditions then forbid a negative cosmological constant on the brane-world.

\bs\bs

\nt{\bf 1. \ INTRODUCTION}
\ss
Cosmology is a natural arena in which to confront string theory with observations.  There are of course many technical questions, mainly connected with the early Universe, which the theory might be expected to address; but in the background there are two very basic, apparently unrelated obstructions to applying string theory to cosmology.  The first is the familiar fact that string and $M$ theory require many more space-time dimensions than we actually see.  It is not enough merely to show that there are space-time structures that are {\it consistent} with the seeming low-dimensionality of the world : we must strive to {\it explain} this fundamental observation.  Secondly, recent observations strongly suggest (see [1] [2] [3] [4] for a selection of references and theoretical perspectives) that the cosmological constant of the observed Universe is small but positive.  From a string-theoretic point of view, this is not a very welcome development [5] since, in the memorable phrase of [6], of all the things one can get out of string theory, a positive cosmological constant does not seem to be one of them.

\ss

We wish to argue that these two problems are not entirely unrelated, and that they can be understood within a single framework.  That framework is the study of the relationship between the stability of matter fields in string theory and the topology of space-time [7].  Specifically, we show that a very simple modification of the topology of the AdS/CFT correspondence [8] has the consequence that certain bulk matter configurations can be stable only if there is a ``brane-world'' [9] in the bulk, and only if the cosmological constant of this brane-world is not negative.

\ss

The idea that the brane-world is the right way to mediate between string theory and cosmology is supported by at least two pieces of evidence.  The first is the well-known fact that one can naturally arrange for the inhabitants of a brane-world to gain the impression that the Universe has fewer dimensions than is actually the case $-$ the brane-world is an ``alternative to compactification''.  On the other hand, it is now generally accepted that the brane-world can be related to string theory through  a ``complementarity'' with AdS/CFT [10] [11] [12] [13] [14].  As the brane-world space-time structure is postulated rather than derived, this does not {\it explain} why the Universe seems not to be ten or eleven dimensional, but it does show us the direction in which such an explanation must be sought.  The second piece of evidence that this is the correct approach is the fact that brane-worlds naturally involve {\it warped product} metrics.  An ordinary Riemannian product of an Einstein manifold of positive scalar curvature with another Riemannian manifold cannot be an Einstein manifold of negative scalar curvature; but a warped product can easily perform this feat (see [15], page 267).  Again, we do not have an {\it explanation} here, but a brane-world embedded in a warped ``bulk'' does supply a very natural way of reconciling an observed positive cosmological constant (on the brane) with a fundamental negative cosmological constant (in the bulk).

\ss

In fact, there is a sense in which the Randall-Sundrum alternative to compactification is much more natural than ordinary compactification itself.  Our basic problem is this : evidently, according to string theory, the observed Universe is a special, privileged submanifold of the full, higher-dimensional world.  What distinguishes this particular submanifold from all the others?  Randall and Sundrum answer : the distinction is {\it geometric}; our Universe is the locus along which the metric of the bulk fails to be smooth.  The sharp ``kink'' in the Randall-Sundrum metric,
$$g^{RS} = dy \otimes dy + e^{-2|y|/L} \; g^F_{ij} \; dx^i \otimes dx^j, \eqno{(1.1)}$$

\nt (where $g^F$ is flat) should therefore be regarded as a virtue, not a vice. It has, however, occasioned some disquiet, and the hope has been expressed [16] that it might ultimately be possible to replace it with something smoother.  This has proved difficult [17] [18] [19], and well-founded scepticism was expressed in [20] as to whether it is possible.  Even the much less welcome negative-tension branes seem to resist smoothing : one can replace them by a ``bounce'' in the warp factor [21], [22], but only at the cost of a negative cosmological constant on the brane-world.  Assuming a non-negative cosmological constant on the brane-world, and the presence of scalar matter in the bulk, the authors of [23] gave an elementary proof that a completely smooth warped-product bulk is not possible.  Nor can one hope to do away with negative-tension branes.  As stated above, the first of these results is welcome, since the presence of a non-smooth ``fracture'' in the bulk provides a natural way of distinguishing a particular submanifold, and so points the way to an ultimate explanation of the fact that the Universe is seemingly low-dimensional.  (We shall call such a fracture a ``Randall-Sundrum structure''.)  The second result is acceptable if the negative-tension branes can be kept out of sight.  There is a natural way of keeping exotic matter out of sight : one puts it at the throat or throats of a {\it wormhole} [24], [14].

\ss

In this work, we shall begin in Section 2 with a discussion of the geometry of the AdS/CFT correspondence (in the form used in [12]) designed to motivate the idea (foreseen in a general way in [25]) that the bulk can be topologically non-trivial.  We shall assume, in Section 3, that the bulk develops the simplest kind of wormhole, obtained by attaching a simple ``handle'', and we assume that there is a massive supergravity scalar field in the bulk [20].  We do not assume that there is a Randall-Sundrum structure in the bulk, nor that the metric is everywhere a warped product.  (In order to perform a concrete calculation, we do assume that the wormhole throats are ``long and narrow'', but we do not believe that this is really essential.)  We then invoke methods of global differential geometry to prove that the scalar matter is (infrared) unstable unless the bulk is geodesically incomplete, and we argue that (in this context) incompleteness signals the presence of a Randall-Sundrum structure.  Finally, having established the existence of a brane-world, we shall argue in Section 4 that a different (ultraviolet) stability condition for matter in the bulk requires $-$ again, because of the wormhole topology $-$ the brane-world cosmological constant to be non-negative.  Since the bulk cosmological constant is negative in this model, this in turn means that the bulk metric resembles a warped product.  In short, it may be possible to {\it derive the characteristic features of the brane-world theory from stability criteria and non-trivial space-time topology}.  This is the main point we wish to make in this work; we believe that its validity extends far beyond the highly suggestive but undoubtedly highly over-simplified special constructions we shall consider.

\bs

\nt{\bf 2. \ GENERIC WARPED GEOMETRIES AND FRACTURED WORMHOLES}
\ss

The underlying geometry of the ``Brane New World'' of [12] is a member of a class of geometries which can be described as follows.  Let $z$ be a coordinate taking values in the open interval $(0,2K)$, and let $P^n$ be an $n$-dimensional compact Riemannian Einstein manifold with metric $g^P$ of positive cosmological constant $(n-1)/L^2$.  (Henceforth, unless we state otherwise, we follow [25] and use Euclidean signature.)  Then the generalised Brane New World metric is defined by
$$\eqalignno{
g^{GBNW} &= dz \otimes dz + \sinh^2 ({z\over L}) g^P_{ij} \; dx^i \otimes dx^j, \ z \in (0,K] &\cr
&= dz \otimes dz + \sinh^2 ({2K-z \over L}) g^P_{ij} \; dx^i \otimes dx^j, \ z \in [K,2K). &{(2.1)}\cr}$$

\nt This metric is continuous everywhere, and smooth everywhere except at $z = K$, where there is a Randall-Sundrum structure, that is, a submanifold of $(0,2K) \times P^n$ which is distinguished by the failure of the metric to be smooth there.  The Ricci tensor of the metric $g^{GBNW}$ satisfies
$${\hbox{Ric}}(g^{GBNW}) = -{n\over L^2} \; g^{GBNW} \eqno{(2.2)}$$

\nt for {\it any} $g^P$ satisfying (see [15], page 267)
$${\hbox{Ric}}(g^P) = +{(n-1) \over L^2} \; g^P. \eqno{(2.3)}$$

\nt The metric induced on the brane-world at $z=K$ is $\sinh^2({K\over L}) g^P$, which satisfies
$${\hbox{Ric}}(\sinh^2({K\over L}) g^P) = {n-1 \over L^2 \sinh^2({K\over L})}(\sinh^2({K\over L})g^P). \eqno{(2.4)}$$

\nt Thus we see that a warped product can naturally reconcile a large negative cosmological constant $-{n\over L^2}$ in the bulk with a {\it small}, {\it positive} cosmological constant ${n-1 \over L^2 \sinh^2({K\over L})}$ on the brane-world : we just have to suppose that $K$, the distance from the brane-world to $z=0$ (or $z=2K$), is sufficiently large.  The original Brane New world metric is obtained by taking $P^n$ to be the sphere $S^n$, and $g^P$ to be the ``round'' metric.  In that case it is possible to extend the metric to $z=0$ and $z=2K$, and the topology of the whole space becomes that of the sphere $S^{n+1}$.  (Note that $(0,2K) \times S^n$ is obtained by deleting the ``North and South'' poles of $S^{n+1}$.)   Notice that, in all cases (whether the topology be $(0,2K) \times S^n$ or $S^{n+1}$), $g^{GBNW}$ is geodesically incomplete at $z=K$, simply because the coefficients in the geodesic equation involve derivatives of the metric components.  This is, however, an extremely mild form of geodesic incompleteness : it just signals the presence of the brane-world, and it is certainly not to be compared with geodesic incompleteness of the kind associated with naked singularities.

\ss

We saw that, when $P^n$ is {\it perfectly} spherical, the points $z=0$ and $z=2K$ can be included in the bulk.  But what happens if the metric on $P^n$ is slightly perturbed?  To answer this, it is best to turn to the full space from which the generalised Brane New World was constructed, namely ``Euclidean Anti-de Sitter space'', which is just the hyperbolic space with metric $g^H$ given by
$$g^H = dr \otimes dr + \sinh^2({r\over L}) \; g^S_{ij} \; dx^i \otimes dx^j, \ r \in [0,\infty), \eqno{(2.5)}$$

\nt where $g^S$ is the round metric on the $n$-sphere of radius $L$.  It is helpful to change the coordinate $r$ to an angular coordinate $\theta$, defined by
$$\cot ({\theta \over 2}) = \sinh({r\over L}), \eqno{(2.6)}$$

\nt so that $\theta = \pi$ corresponds to $r = 0$ and $\theta \to 0$ as $r \to \infty$.  The metric is now
$$g^H = {\hbox{cosec}}^2 ({\theta \over 2}) [{1\over 4} \; d\theta \otimes d\theta + \cos^2 ({\theta \over 2}) g^S_{ij} \; dx^i \otimes dx^j]. \eqno{(2.7)}$$

\nt Now clearly 0 to $\pi$ is not a natural range for an angular coordinate : we should allow $\theta$ to run from $-\pi$ to $\pi$.  Doing this means taking two copies of the hyperbolic space and fusing them at $r=0$ ($\theta = \pi$ or $-\pi$) and ``$r=\infty$'' ($\theta = 0$).  Conformal infinity is no longer a {\it boundary}; it has become a {\it submanifold} of the compact manifold $S^1 \times S^n$, where $S^1$ is the circle parameterised by $\theta$.  There are, in fact, technical advantages in this way of thinking about conformal infinity, since it is more general [26].  Notice that the two halves of this space cannot communicate, since $\theta = 0$ is infinitely far away, and $\theta = \pi$ or $-\pi$ is ``choked off'' by the vanishing of $\cos({\theta \over 2})$ there.

\ss

Now the generalised Brane New World metric is obtained by ``cutting and pasting'' the space $S^1 \times P^n$ with (``collapsed wormhole'') metric
$$g^{CW} = {\hbox{cosec}}^2({\theta \over 2})[{1\over 4}L^2d\theta \otimes d\theta + \cos^2({\theta \over 2})g^P_{ij} \; dx^i \otimes dx^j]. \eqno{(2.8)}$$

\nt Again, if $g^P$ is {\it any} Einstein metric with cosmological constant ${(n-1)\over L^2}$, then $g^{CW}$ is Einstein with cosmological constant $-{n \over L^2}$, and of course we regain our ``doubled'' Euclidean AdS space if $P^n = S^n$.

\ss

Our earlier question leading into this digression was : what happens at $\theta = \pi$ and $-\pi$ if $P^n$ is a slightly perturbed sphere?  Let Greek superscripts and subscripts run from 1 to $n+1$, let $R^{CW}_{\alpha\beta\gamma\delta}$ denote the components of the curvature tensor of $g^{CW}$, let $R^P_{ijk\ell}$ be the curvature components for $g^P$, and let $R^P$ be the scalar curvature of $g^P$.  Then we have
$$\eqalignno{
R^{CW}_{\alpha\beta\gamma\delta} R^{CW\alpha\beta\gamma\delta} = &{2n(n+1)\over L^4} 
+\tan^4({\theta \over 2})&\cr
&\ \times [R^P_{ijk\ell}R^{Pijk\ell} 
- {2n(n-1)\over L^4} - {4\over L^2} {\hbox{cosec}}^2({\theta \over 2})\{R^P - {n(n-1)\over L^2}\}]. \ \ \ \ \ \ \ &{(2.9)}\cr}$$

If $P^n$ is perfectly spherical, then
$$R^P_{ijk\ell} = {1\over L^2} (g^P_{ik}g^P_{j\ell} - g^P_{i\ell}g^P_{jk}), \eqno{(2.10)}$$

\nt and so
$$R^P_{ijk\ell} R^{Pijk\ell} = {2n(n-1) \over L^4}, \quad R^P = {n(n-1) \over L^2}, \eqno{(2.11)}$$

\nt so that, in this case, the coefficient of $\tan^4({\theta \over 2})$ in (2.9) is precisely zero.  But if $g^P$ is perturbed, {\it no matter how slightly}, then $R^{CW}_{\alpha\beta\gamma_\delta} R^{CW\alpha\beta\gamma\delta}$ diverges as $\theta \to \pm \pi$.  It follows that the generalised Brane New World manifold, with metric (2.1), will develop curvature singularities as soon as $P^n$ is perturbed, and this is true even if $g^P$ and $g^{GBNW}$ are Einstein metrics (because, in that case, the second equation of (2.11) still holds but the first does not, in general).  For any $g^P$, the metric (2.8) clearly represents an infinite space with ``infinity'' at $\theta = 0$; the metric induces a conformal structure there, represented by $g^P$.  The bulk has a ``collapsed wormhole'' at $\theta = \pm \pi$, since the metric pinches off there.  What we have found is that this pinching-off is non-singular only in the extremely special case where $P^n$ is {\it exactly} spherical.  The slightest deviation from perfect spherical symmetry causes the space to become violently singular in the region around the point where the wormhole has collapsed.

\ss

String theory has taught us [27][28] that singularities of this kind are not necessarily an indication that a theory is pathological.  In the brane-world context, in which the bulk is taken to be very ``real'' (as opposed to an interpretation of AdS/CFT in which the correspondence is regarded more formally [29]), this means that these singularities are resolved in some way.  The natural and obvious way is to {\it open up the wormhole throats} : we regard the collapsed wormhole metric, (2.8), as the (generically) singular limit of a non-singular open wormhole metric.  In fact, the necessity of resolving singularities of this kind has been discussed in the brane-world literature, but usually in the context of bulk metrics of the form
$${L^2\over z^2} (dz \otimes dz + g^{RF}), \eqno{(2.12)}$$

\nt where $z \in (0,\infty)$ and $g^{RF}$ is Ricci-flat.  This metric is Einstein with negative cosmological constant $-{n\over L^2}$ ([15], page 267), and it is non-singular as $z\to\infty$ only if $g^{RF}$ is actually {\it flat} [30].  When $g^{RF}$ represents a black hole, we therefore seem to have a black string extending to a nakedly singular AdS horizon.  It was argued in [31] that this is an unstable configuration and that Gregory-Laflamme instability [32] will lead to the singular horizon being resolved in some way.  In another approach [33] [34] it is argued that localised matter on or near the brane-world can only generate localised gravitational fields which (when higher Randall-Sundrum modes are taken into account) decay rapidly towards the horizon, so that the horizon geometry will be unchanged in a physically realistic scenario (see also [35], [36]).  This works well for the metric (2.12), wherein the singularity is infinitely far off; but it is much less convincing in the case where the brane-world has a positive cosmological constant, since (2.8) clearly indicates that the singularity (at $\theta = \pi$) is {\it not} infinitely remote in this case.  (This is an interesting example of a situation in which there is a demonstrable difference between a positive cosmological constant $-$ no matter how small $-$ and a cosmological constant which is precisely zero.)

\ss

All this supports the contention that the sharp ``cusp'' in (2.8) is physically unrealistic: it must be replaced by some structure which does not instantly become singular when $P^n$ ceases to be perfectly spherical.  We argued above that the obvious course of action is to ``open up the wormhole''.  It is always possible to do this by excising the collapsed section; the only difficulty is to decide what to insert in its place.  The simplest possible choice is to perform a ``graft'', as for example in the case of the following ``open wormhole" metric :
$$\eqalignno{
g^{OW}
&={\hbox{cosec}}^2({\theta\over 2})[{1\over 4}L^2d\theta \otimes d\theta +  \cos^2({\theta\over 2})g^P], \ \theta \in (0,\alpha] \cup [-\alpha,0) &\cr
&={\hbox{cosec}}^2(\alpha-{\theta\over 2})[{1\over 4}L^2d\theta \otimes d\theta +  \cos^2(\alpha -{\theta\over 2})g^P], \ \theta \in [\alpha,\pi] &\cr
&={\hbox{cosec}}^2(\alpha + {\theta\over 2})[{1\over 4} L^2d\theta \otimes d\theta + \cos^2(
\alpha+{\theta \over 2})g^P], \ \theta \in (-\pi,-\alpha]. &(2.13) \cr}$$

\nt Here $\alpha$ is a constant angle in $({\pi\over 2}, \pi)$.  The collapsed wormhole metric (2.8) is obtained in the limit $\alpha \to \pi$; otherwise $\alpha$ and $-\alpha$ represent the throats of the wormhole.  Deep inside, at $\theta = \pi$, there is a Randall-Sundrum structure corresponding to a positive-tension brane-world.  The metric of the brane-world is given by
$$g^{BW} = \tan^2(\alpha)g^P. \eqno{(2.14)}$$

\nt If $g^P$ is an Einstein metric with cosmological constant ${(n-1)\over L^2}$, then $g^{OW}$ is Einstein with cosmological constant $-{n\over L^2}$, and $g^{BW}$ is Einstein with cosmological constant
$$\Lambda^{BW} = {+(n-1)\over L^2\tan^2(\alpha)}. \eqno{(2.15)}$$

\nt Near to the brane-world, this is of course just the generalised Brane New World in a different coordinate system.  Globally, however, the manifold is quite different : instead of the delicate, generically singular cusps at $z=0$ and $z=2K$ in (2.1), we have wormhole throats at $\theta = \pm \alpha$, leading to an infinite Euclidean AdS region on the other side, with conformal infinity at $\theta = 0$.

\ss

Since $\Lambda^{BW}$ is observed to be very small, (2.15) tells us that the angle $\alpha$ must be only slightly larger than ${\pi \over 2}$.  The metric of the wormhole throats, $g^{WT}$, is
$$g^{WT} = \cot^2({\alpha \over 2}) \cot^2(\alpha)g^{BW}, \eqno{(2.16)}$$

\nt which shows that the extreme smallness of the observed cosmological constant is just a reflection of the fact that the wormhole throats are extremely narrow (and far from the brane-world).

\ss

Clearly, this wormhole structure provides a concrete way of formulating the complementarity [13] between AdS/CFT and the Randall-Sundrum theory, as well as a natural way of avoiding naked singularities in the bulk.  Physically, one can imagine beginning with an AdS-like bulk (arising in the string context in the guise of AdS/CFT) in which a wormhole develops.  We have argued that it is natural to suppose that the brane-world inhabits the wormhole.  The real challenge, however, is to prove that such a wormhole {\it must} contain a brane-world, and that the latter should have a positive cosmological constant.  We do not claim to have such a proof, but in the next section we shall uncover strong evidence that these statements are valid.

\bs

\nt{\bf 3. \ NON-SMOOTHABLE WORMHOLES IN AdS/CFT}.
\ss
We begin by formulating AdS/CFT for the simplest possible wormhole topology.  Let $P^n$ be a compact $n$-dimensional manifold, and set ${\widehat W}^{n+1} = S^1 \times P^n$, where $S^1$ is parametrised by an angle $\theta \in (-\pi, \pi]$.  Let $g^W$ be a piecewise smooth Riemannian metric on
$$W^{n+1} = \{ S^1 - (\theta = 0)\} \times P^n \eqno{(3.1)}$$

\nt such that it is possible to find a piecewise smooth function $F$ on ${\widehat W}^{n+1}$ with the following properties :

\i (i)/ $F=0$ if and only if $\theta = 0$;

\i (ii)/ $F$ is smooth at $\theta =0$, and $dF \not= 0$ there;

\i (iii)/ $F^2g^W$ extends continuously to a metric on all of ${\widehat W}^{n+1}$;

\i (iv)/ Let $|dF|_F$ be the norm of $dF$ with respect to the extension of $F^2g^W$.  We require that this norm, evaluated at $\theta = 0$, should not depend on position in $P^n$.

\ss

\nt The first three conditions just mean that the submanifold at $\theta = 0$ (which is diffeomorphic to $P^n$) is infinitely far from all points in the ``bulk''.  The last point ensures that all sectional curvatures along geodesics which ``tend to infinity'' shall approach a common negative value; that is, it ensures that the geometry ``near infinity'' resembles (Euclidean) AdS$_{n+1}$.  (See [37] and references therein.)  Of course, AdS/CFT in this context claims that a certain conformal field theory at $\theta = 0$ is dual to a gravitational theory on $\theta \not= 0$; the latter space resembles AdS$_{n+1}$ near $\theta = 0$, but there is a wormhole deeper in the bulk.

\ss

Now we wish to ask : can a ``physically reasonable'' AdS/CFT wormhole be smooth?  In order to understand some of the subtleties we shall encounter, let us consider some simple examples.  The collapsed wormhole metric (2.8), is generically singular because of the term involving $\cos^2({\theta\over 2})$.  If we drop that coefficient, we obtain a ``smooth open wormhole'' metric which is indeed perfectly smooth everywhere (even at $\theta = \pm \pi$) :
$$g^{SOW} = {\hbox{cosec}}^2({\theta \over 2}) [{1\over 4} L^2 d\theta \otimes d\theta + g^P_{ij} \; dx^i \otimes dx^j]. \eqno{(3.2)}$$

\nt Note that this satisfies all four conditions stipulated above, with $F = \sin({\theta \over 2})$; the asymptotic value of all sectional curvatures is $-{1\over L^2}$, for all $g^P$.  However, the non-zero (1,1) components of the Ricci tensor are, in an obvious notation,
$$\left.
\eqalign{
&(R^{SOW})^\theta_\theta = -{n\over L^2}. \cr
&(R^{SOW})^i_j = -{n\over L^2} \; \delta^i_j + \sin^2({\theta \over 2})\left[(R^P)^i_j + {(n-1)\over L^2} \; \delta^i_j\right].\cr} 
\quad \right\}\eqno{(3.3)}$$

\nt Thus we see that the bulk can be an Einstein manifold (or a slightly perturbed Einstein manifold) only if $P^n$ is (approximately) an Einstein manifold of {\it negative} scalar curvature.  Now clearly the metric (3.2) induces a conformal structure at infinity represented by $g^P$, so we conclude that the bulk is Einstein if and only if the conformal structure at infinity is represented by a metric of negative scalar curvature.  In that case, however, it is known that string theory in the bulk is {\it unstable} to the emission of ``large'' branes [38], [39].  This is our first indication that, in string theory, it may be difficult to reconcile smoothness, stability, and physically reasonable matter on a topologically non-trivial background.

\ss

As a second attempt, we can enforce stability by requiring $P^n$ to have positive scalar curvature.  If we take it to be Einstein with cosmological constant ${(n-1)\over L^2}$, then the Ricci components of the bulk are
$$\left.
\eqalign{
&(R^{SOW})^\theta_\theta = -{n\over L^2} \cr
&(R^{SOW})^i_j = \left[-{n\over L^2} + {2(n-1)\over L^2} \sin^2({\theta\over 2})\right]\delta^i_j \cr}
\quad \right\} \eqno{(3.4)}$$

\nt Thus we see that some kind of matter is present.  Clearly the eigenvalues of the Ricci tensor all approach $-{n\over L^2}$ as infinity $(\theta =0$) is approached, but this is not enough to ensure physically reasonable behaviour.  The definition of an asymptotically AdS spacetime [40] imposes conditions on the rate at which the stress-energy-momentum tensor should decay as infinity is approached.  These can be interpreted as conditions on the functions $\lambda_\beta$, the eigenvalues of the (1,1) version of the Ricci tensor.  In [40] it is pointed out that (in our notation) a physically reasonable condition, in the case of a four-dimensional spacetime which is asymptotically locally a space of constant sectional curvature $-{1\over L^2}$, is that the $\lambda_\beta$ should all satisfy
$$F^{-4}(\lambda_\beta +{n\over L^2}) \to 0 \quad {\hbox{as}} \quad F \to 0. \eqno{(3.5)}$$

\nt (Here it is reasonable to assume uniform convergence.  In [40], $F^{-3}$ is actually used, but the authors mention that $F^{-4}$ is justified physically. In fact, since we are interested in spacetimes which are at least five-dimensional, even stronger fall-off conditions could be justified.) In the present case, $F = \sin({\theta \over 2})$, and it is evident that (3.5) is {\it not} satisfied.  Thus, instability can be averted only at the cost of introducing physically unreasonable matter fields.

\ss

Both of these problems can be solved by dropping the requirement of smoothness : for if the metric $g^P$ satisfies (2.3), then (2.13) induces at $\theta =0$ a conformal structure represented by a metric of positive scalar curvature, and furthermore (3.5) is trivially satisfied.  One begins to suspect that stability and appropriate asymptotic behaviour might {\it force} the wormhole metric to be non-smooth.  As a non-smooth wormhole can always, by suitable ``cutting and pasting'', be interpreted as one which (like the wormhole represented by (2.13)) contains at least one brane-world, we would be justified in interpreting such a result as implying that {\it brane-worlds inevitably arise when} AdS/CFT {\it is formulated on a space with the wormhole topology}.  In order to discuss this in a concrete way, we need to introduce a theorem due to Witten and Yau [39] and improved by Cai and Galloway [41].  Recall first (see, for example, [25]) that a Riemannian manifold is said to have a conformal compactification if it can be regarded as the interior of a compact manifold-with-boundary, such that the boundary is the zero locus of a function $F$ which satisfies four conditions analogous to those stated at the beginning of this section.  Note that there is no reason, in general, to expect the conformal boundary to be connected, even if the interior is connected.  Then the relevant part of the Witten-Yau-Cai-Galloway theorem may be stated as follows.

\ms

\nt{\bf THEOREM 1} : Let $M^{n+1}$ be a complete, connected, $(n+1)-$dimensional Riemannian manifold which admits a conformal compactification with boundary $N^n$.  Suppose that the following three conditions hold :

\i (a)/ The eigenvalue functions of the Ricci tensor of $M^{n+1}$ all satisfy $\lambda_\beta \ge -{n\over L^2}$;

\i (b)/ The $\lambda_\beta$ all approach $-{n\over L^2}$ towards infinity, in accordance with (3.5) discussed earlier.

\i (c)/ $N^n$ has at least one connected component such that the conformal structure induced there is represented by a metric of non-negative scalar curvature.

\ss

Then $N^n$ must be connected.

\ss

This result has the following consequence.

\ss

\nt{\bf THEOREM 2} : Let $P^n$ be a compact, connected, $n-$dimensional manifold, and let ${\widehat W}^{n+1}$, $W^{n+1}$, and $g^W$ be as described at the beginning of this section.  Suppose that the wormhole is open and that $g^W$ satisfies the following conditions :

\i (a)/ The eigenvalue functions of the Ricci tensor of $g^W$ all satisfy $\lambda_\beta \ge -{n \over L^2}$;

\i (b)/ The $\lambda_\beta$ all approach $-{n \over L^2}$ as $\theta \to 0$, in accordance with condition (3.5) above;

\i (c)/ The conformal structure induced at $\theta = 0$ is represented by a metric of non-negative scalar curvature.

\ss

Then $W^{n+1}$ is {\it not} geodesically complete with respect to $g^W$.

\ss

\nt{\bf PROOF} : Instead of letting $\theta$ run from $-\pi$ to $+\pi$, let it run from 0 to $2\pi$.  Then, since the wormhole is open, $W^{n+1}$ is $(0,2\pi) \times P^n$, which is the interior of the compact manifold-with-boundary $[0,2\pi] \times P^n$.  Since $(0,2\pi) \times P^n$ is connected, all of the conditions of Theorem 1, with the possible exception of completeness, are satisfied.  But the boundary of $[0,2\pi] \times P^n$ is not connected; thus, in fact, $g^W$ cannot be complete.  This concludes the proof.

\ss

Notice here the crucial role played by the assumption that the wormhole is {\it open} : we need this in order to ensure that $W^{n+1}$ is {\it connected}, as required by Theorem 1.  If the wormhole has collapsed, then $W^{n+1}$ is disconnected, and Theorem 2 can fail.  For example, the metric is (2.8) is geodesically complete if $P^n$ is $S^n$ with the round metric.  Theorem 2 states that opening up the wormhole inevitably leads to incompleteness, as in (2.13), provided that the three conditions are satisfied (as they are by (2.13), if $g^P$ satisfies (2.3).)  We now consider the physical meaning of those conditions.

\ss

Condition (c) is, as we have seen, a stability condition : without it, string theory on $W^{n+1}$ will be unstable ``near'' the conformal ``boundary''.  (Note that this is a ``UV'' instability in the sense of [42].)  Next, recall that the eigenvalues $\lambda_\beta$ deviate from $-{n\over L^2}$ (the cosmological constant of the bulk) only if matter is present in the bulk, and we saw that condition (b) just means that this matter should behave at large distances in a way that is consistent with the asymptotically hyperbolic metric.  That is, (b) is a physical condition, which requires the matter to behave in a reasonable way that is compatible with the assumed form of the asymptotic geometry.

\ss

Condition (a) is new, and it requires the most care in interpretation.  It is tempting to assume that it is the Euclidean analogue of the Strong Energy Condition, familiar from the singularity theorems [43], {\it but that is not correct}.  The relationship between the metric and the Ricci tensor is highly complex, and we cannot assume that a Lorentzian metric which satisfies the SEC will, when ``Euclideanized'', satisfy condition (a).  In fact, the reverse is the case.  Consider the case of a scalar field $\phi$ with a potential $V(\phi)$ (which does {\it not} include the cosmological constant) and a stress-energy-momentum tensor with components
$$T_{\mu\nu} = \partial_\mu \phi \; \partial_\nu\phi - {1\over 2} g_{\mu\nu} (\partial_\alpha\phi \; \partial^\alpha\phi) - g_{\mu\nu}\; V(\phi). \eqno{(3.6)}$$

\nt (We use the ``mostly plus'' signature in the Lorentzian case.)  Then, on an $(n+1)-$dimensional manifold with a background cosmological constant $-{n\over L^2}$, we have
$$R_{\mu\nu} + {n\over L^2} g_{\mu\nu} = \partial_\mu \phi \; \partial_\nu \phi + {2\over n-1} \; g_{\mu\nu} \; V(\phi). \eqno{(3.7)}$$

\nt Now in the Lorentzian case, a unit future-pointing time like vector field with components $t^{\mu}$ satisfies $g_{\mu\nu} t^{\mu} t^{\nu} = -1$, so a {\it positive} $V(\phi)$ will {\it reduce} $R_{\mu\nu}t^{\mu}t^{\nu}$, leading to possible violations of the Strong Energy Condition.  But in the Euclidean regime, if $t^{\mu}$ are the components of a unit eigenvector of the Ricci tensor, corresponding to an eigenvalue function $\lambda$, then we have
$$\lambda + {n\over L^2} = (t^{\mu} \partial_\mu \phi)^2 + {2\over n-1} \; V(\phi), \eqno{(3.8)}$$

\nt and so the positivity of $V(\phi)$ is just what is needed to see to it that condition (a) in Theorem 2 is satisfied.  Far from ensuring that the Strong Energy Condition is satisfied, then, condition (a) requires the presence of matter that tends to violate it.

\ss

More generally, if we have bulk matter in $n+1$ dimensions with a diagonalized stress-energy-momentum tensor
$$T_{\mu\nu} = {\hbox{diag}} (\rho, p_1, p_2, ..., p_n), \eqno{(3.9)}$$

\nt then Einstein's equation yields Ricci components (with respect to an orthonormal basis)
$$\eqalignno{
&R_{oo} = -{n\epsilon \over L^2} + ({n-2\over n-1})\rho - {\epsilon \over n-1} \; \sum^n_{j=1} \; p_j &{(3.10)} \cr
&R_{ii} = -{n\over L^2} - {\epsilon \rho \over n-1} + p_i - {1\over n-1} \; \sum^n_{j=1} \; p_j, &{(3.11)}\cr}$$

\nt where a zero subscript refers to time in the Lorentzian case, and where $\epsilon = 1$ in the Euclidean case, and $\epsilon = -1$ in the Lorentzian case.  For the Euclidean case, summing on $i$ in equation (3.11) yields, for the Ricci eigenvalues $\lambda_i = R_{ii}$,
$$\sum^n_{i=1} (\lambda_i + {n\over L^2}) = -{(n\rho + {\mathop{\sum}\limits^n_{i=1}}\; p_i) \over n-1}. \eqno{(3.12)}$$

\nt But in the Lorentzian case, we have
$$\sum^n_{i=1} \; (R_{oo} + R_{ii}) = n\rho + \sum^n_{i=1} \; p_i. \eqno{(3.13)}$$

\nt Clearly, (3.12) and condition (a) in Theorem 2 require the right side of (3.13) to be non-positive.  But the Strong Energy Condition requires $R_{\mu\nu}t^\mu t^\nu \ge 0$ for all timelike vectors, so it demands (among other conditions) that each term on the left side of (3.13) be non-negative.  Thus condition (a) requires that the Lorentzian version of the theory should {\it violate} the SEC if there is matter in the bulk (other than the cosmological
constant itself, for which the right sides of (3.12) and (3.13) are both zero).

\ss

This conclusion may seem strange, but in fact it is to be expected.  For it is well known [24] that {\it Lorentzian wormholes can only be sustained by matter which violates energy conditions} (typically the Null Energy Condition or its ``averaged'' version, violation of which implies a violation of the SEC).  As is emphasised in [44], in theories (like string theory) involving scalar fields, such matter is abundant and should not be considered particularly outlandish.  Notice that even our simplest wormhole, with metric (2.13), contains ``exotic'' matter in the form of negative-tension branes at $\theta = \pm \alpha$.  The prevalence of negative-tension branes in brane-world theories (as part of the {\it background}, in which role they are harmless [16]) is now understandable and natural : one expects such objects in a wormhole.

\ss

In short, then, condition (a) in Theorem 2 is another physically well-motivated requirement : it just reminds us that if we wish to have a wormhole in the AdS/CFT bulk, then we should see to it that we have the right kind of matter to sustain such a structure.  Summarizing, Theorem 2 has the following physical interpretation.  Suppose that we wish to do string theory on an asymptotically (Euclidean) AdS- like manifold which contains a wormhole deep in the interior.  Suppose further that the matter content of the bulk is (a) able to sustain a wormhole and (b) consistent with the asymptotic AdS geometry, and also that (c) the string theory is UV stable against the emission of large branes.  Then if the bulk contains no singularities, it must contain a Randall-Sundrum structure, as in (2.13).

\ss

This result means that, if we can plausibly argue that string theory generates ``wormhole matter'', then we need no longer {\it postulate} the existence of ``fractures'' in high-dimensional space-times, on which gravity is localised, inducing the illusion of low-dimensionality $-$ instead, we can {\it deduce} it.  We now argue that this may well be the case.

\ss

It was shown in [20] (see [45] for a summary, and [46] for further developments) that brane-worlds can be realised in a supersymmetric context by means of a careful analysis of massive supergravity scalars in the bulk.  (The throats of the wormhole correspond to the infrared in that context.)  Following [20], we consider a supergravity scalar $\phi$ which approaches a fixed value $\phi_\ast$ at a critical point of the potential.  Keeping the negative background cosmological constant separate, as above, we thus have a potential, to lowest non-trivial order, of the form
$$V(\phi) = {1\over 2} m^2 (\phi - \phi_\ast)^2. \eqno{(3.14)}$$

\nt It may seem that we are assuming here what we seek to prove, namely that $V(\phi)$ is positive.  We must bear in mind, however, that on a background with a negative cosmological constant, there is no guarantee that $m^2 > 0$; all we have is the Breitenlohner-Freedman bound [47] which requires $m^2$ to be greater than a fixed negative value.  On the other hand, the fact that $m^2 < 0$ is permitted in general does not mean that it is allowed in this particular case.

\ss

As we saw earlier, equation (2.15) for the brane-world cosmological constant, and equation (2.16) for the wormhole throat metric, indicate that the throats are narrow and far away from the brane-world (in that model of the wormhole).  It is reasonable to suppose that this is generic for a wormhole consistent with the observed very small value of the cosmological constant.  In other words, we assume that the wormhole dynamics is such that the regions around the throats are ``long and narrow''.  In such a geometry, transverse variations in the shape of the cross-sections are not important, so we shall assume that the metric near the throats can be {\it approximated} by a warped product of the local form
$$dy \otimes dy + e^{2A(y)} \delta_{ij} \; dx^i \otimes dx^j, \eqno{(3.15)}$$

\nt where we follow the notation of [20]; here $y$ is a parameter measuring length along the long, narrow part of the wormhole, so it becomes very large towards the throats.  (We stress that we do {\it not} assume that (3.15) is valid, even approximately, away from this part of the wormhole.)  Clearly $A(y)$ must tend to $-\infty$ as $y$ becomes large.

\ss

Now we introduce the massive super gravity scalar field $\phi$ discussed in [20], with a stress-energy-momentum tensor given by (3.6) and a potential given by (3.14).  Ignoring any transverse variations in $\phi$, we can solve the equation of motion and obtain [20] the following solutions,for a four-dimensional brane-world: 
$$\phi = \phi_\ast + ce^{-E_oA(y)}, \ \phi = \phi_\ast + c\; e^{-(4-E_o)A(y)}, \eqno{(3.16)}$$

\nt where $c$ is a constant and $E_o$ is the lowest energy in an approximately locally AdS background of curvature $-{1\over L^2}$; to this order of approximation we have from [20]
$$E_o = 2 + \sqrt{m^2L^2+4}. \eqno{(3.17)}$$

\nt The Breitenlohner-Freedman bound here is $E_o \ge 2$, showing indeed that in general we can only require $m^2 \ge -{4\over L^2}$.  But in this particular case, we see that equations (3.16) imply that $\phi$ is unstable towards the IR (that is, as $A(y) \to -\infty$) for $2 \le E_o < 4$.  (Of course, $A(y)$ does not really diverge in the wormhole case, but the failure of $\phi$ to approach $\phi_\ast$ for $E_o$ in this range is still indicative of instability.)  As in [20], we conclude that the IR stability of $\phi$ requires
$$E_o > 4. \eqno{(3.18)}$$

\nt But (3.17) means that this is equivalent to $m^2 > 0$, which, by (3.14), means $V(\phi) \ge 0$.  In turn, (3.8) now ensures that condition (a) of Theorem 2 is satisfied.  Let us assume, as is reasonable, that $\phi$ decays in an acceptable way towards infinity (condition (b)) and that $\phi$ does not distort the geometry in such a way as to induce instability due to emission of large branes (condition (c)).  Then Theorem 2 means that the wormhole must fracture : some structure like the one represented by (2.13) must exist in the wormhole, if a naked singularity is to be avoided.

\ss

In [20], the condition (3.18) is obtained as a {\it necessary} condition which must be satisfied if the massive supergravity scalar is to sustain a brane-world. Here we have argued that the presence of such a scalar field satisfying this condition is also {\it sufficient} to ensure the existence of a Randall-Sundrum structure.  In our view, the real role of the scalar field is to keep the wormhole open.  It can do this provided it satisfies certain stability conditions in both the UV and the IR; the existence of the brane-world is by-product of those conditions.

\ss

This analysis of the way in which scalar matter sustains both the wormhole and the brane-world is obviously crude and approximate.  Nevertheless we believe that it points the way to an explanation of the fact that we appear to live on a low-dimensional fracture in a fundamentally high-dimensional Universe.  We shall now argue that this theory also predicts the correct sign for the observed cosmological constant.

\bs

\nt{\bf 4. WHY THE BRANE-WORLD COSMOLOGICAL CONSTANT CANNOT BE NEGATIVE}
\ss
Let us agree that we have a brane-world in the wormhole; we wish to ask whether any particular sign is favoured for the cosmological constant of the brane-world.

\ss

We have mentioned several times that stability conditions require that the conformal structure at infinity must be represented by a metric of either zero or positive scalar curvature.  However, this does not in itself imply that the scalar curvature of the brane-world satisfies a similar condition.  For example, if $g^{RF}$ in (2.12) is the usual flat metric on $\R^n$, then (2.12) is the canonical hyperbolic metric on $\R^{n+1}$.  The conformal boundary is just the sphere $S^n$ with its usual conformal structure represented by a metric of positive scalar curvature.  Thus there is no correlation between the scalar curvatures of the cross-sections and that of conformal infinity in general, even when the bulk metric is assumed to be a warped product.

\ss

In the case of the wormhole topology $S^1 \times P^n$, the brane-world {\it is} diffeomorphic to the infinity submanifold (as in (2.13)), but this still does not guarantee that the brane-world has a non-negative cosmological constant.  For while we may habitually associate particular topologies with certain signs of the scalar curvature $-$ the sphere $S^n$ with positive, and the torus $T^n$ with zero scalar curvature $-$ this applies only to canonical metrics.  In fact, there exist metrics of {\it constant negative} scalar curvature on $S^n$ and $T^n$ for all $n \ge 3$ ([15], page 123).  Now the conditions assumed in Theorem 2 allow a wide choice of bulk metrics; surely they do not guarantee that all of the cross-sections must have similar geometries.  One can certainly imagine a bulk geometry which interpolates, for example, between a brane-world with topology $S^4$, and {\it negative} scalar curvature, and an infinity with topology $S^4$ and a conformal structure represented by a metric of {\it positive} scalar curvature.  In general, then, the sign of the scalar curvature at infinity does not tell us very much about the cosmological constant of the brane-world.

\ss

However, for cosmological purposes we are not interested in arbitrary geometries on the brane-world; in this context, we really just want to distinguish between flat, deSitter, and anti-deSitter branes, and here we can say something definite.  First, recall that the Euclidean versions of deSitter and anti-deSitter spaces are, respectively, the spaces of constant positive and negative sectional curvature.  Because, in the wormhole picture, the brane-world has the same topology as infinity, and because it is important [25] that the latter be compact in the AdS/CFT framework, we must compactify.  That is, for flat brane-worlds we take a torus $T^n$ or its quotient $T^n/F$ by a finite group; for deSitter brane-worlds we take $S^n$ or one of its quotients; and for anti-deSitter brane-worlds we take $\R^n/ \Gamma$, where $\Gamma$ is an infinite discrete group such that the quotient is a {\it compact} hyperbolic space.  Now suppose that we have an anti-deSitter brane-world $-$ that is, suppose that the {\it observed} cosmological constant is negative.  Then, since each cross-section of $S^1 \times P^n$ has the same topology, infinity has the topology $\R^n/ \Gamma$.  As we have explained, identical topology does not necessarily entail similar geometry.  But in this particular case, Gromov and Lawson (see [48], page 306) have proved the following remarkable result.

\ms

\nt{\bf THEOREM 3} : A compact manifold which carries a Riemannian metric of {\it sectional}  curvature $\le 0$ (or $< 0$) cannot carry any metric of {\it scalar} curvature $> 0$(respectively, $\ge 0$). 

\ss

\nt This theorem means that for {\it any} conformal structure on $\R^n/ \Gamma$, if we pick a representative metric of constant scalar curvature, then this constant {\it must} be negative.  We shall therefore find instability to the emission of ``large branes'' near infinity, {\it no matter what the bulk metric may be}.  It follows that an anti-deSitter brane-world is ruled out : the observed cosmological constant cannot be negative.  The power of this argument derives from its topological nature.  No matter how we distort the bulk metric, as long as it admits a string-theoretic interpretation (that is, technically, as long as it continues to satisfy the four asymptotic conditions given at the beginning of Section 3) then a negative cosmological constant on the brane will cause instability in the bulk.  String theory may well require a negative cosmological constant in the bulk, as we have assumed : but it forbids a negative cosmological constant on the brane, if a wormhole is present.

\ss

The Gromov-Lawson theorem also indicates that we should not expect the brane-world cosmological constant to be zero.  For the theorem implies that, on a manifold with the topology of a compact flat manifold, it is impossible to construct any Riemannian metric of positive scalar curvature.  Thus, for {\it any} conformal structure on such a manifold, if we pick a representative metric of constant scalar curvature, then this constant will be either zero or negative.  Presumably, therefore, a ``generic'' perturbation of a flat metric produces a metric in a conformal structure represented by a metric of constant negative scalar curvature.  Therefore, while it is true that a brane-world with a vanishing cosmological constant does not induce instability due to emission of ``large branes'' in the bulk, a generic, arbitrarily small perturbation of the brane geometry {\it will} have this effect.  In short, a brane-world with a vanishing cosmological constant is on the brink of causing UV instability, and the Gromov-Lawson theorem implies that perturbations are more likely to push the system over the brink than to keep it there.

\ss

The results of this section can be summarized as follows.  Suppose that, as was argued in Section 3, a brane-world exists in an AdS/CFT wormhole.  Then a combination of stability conditions and global differential-geometric theorems shows that a negative cosmological constant on the brane is ruled out, and the same is probably true of a vanishing cosmological constant.  Essentially, the observed cosmological constant has to be positive so that string theory can be stable. Notice that this conclusion depends on taking into account effects
(the communication of the topology of the brane-world to that of infinity, and instability due
to the emission of large branes) that are very much non-perturbative and "stringy". If one does
not embed the brane-world in string theory, or if non-perturbative effects are not taken into account, then the conclusion is false: for example, in reference [49] and in many more recent works, seemingly consistent supersymmetric anti-deSitter branes are constructed. We are suggesting that in the full, non-perturbative theory, these solutions will not be stable. 

\bs

\nt{\bf 5. CONCLUSION}
\ss
Much concern has recently been expressed (for example, in [50] and [51]) regarding the
difficulty of obtaining a positive cosmological constant in string theory. It is important
to distinguish between two separate kinds of concerns. First, one might wonder whether
it is even {\it possible} to construct a deSitter type Universe in the string context without
generating unphysical singularities in the bulk or other obviously unacceptable behaviour.
This is the question considered in [28]. Only when this question is answered can one go
on to the second question: whether one can reconcile a deSitter Universe with deeper but less
well-understood aspects of string theory, such as the fact that the latter always 
apparently involves an infinite number of degrees of freedom, or any of the other issues
raised in [50] and [51]. 

\ss

Our objective here has been to show how a deSitter Universe can at least be constructed
within a higher-dimensional geometry which certainly is compatible with the principles
of string theory. The basic idea, implicit already in [12], is to assert that the apparent positive cosmological constant is a consequence of the apparently low-dimensional nature of our world.  The real, higher-dimensional world {\it does} have a negative cosmological constant : only our small corner of it goes against the grain.  The problem, of course, is to prove that string theory can naturally  predict such an apparently contrived state of affairs.

\ss

From our point of view, the ``no-go'' theorem of [28] should be interpreted as stating that if the brane-world has a positive cosmological constant (that is, if the Strong Energy Condition is violated on the brane), then the bulk scalar must also violate the Strong Energy Condition.  The problem, then, is to find a context in which such violations are natural or indeed inevitable.  The choice is clear : wormholes require ``exotic'' matter.  Our plan is to make a virtue of that necessity by using wormholes in the AdS/CFT bulk as a controllable way of violating the bulk SEC. We saw that a brane-world with a positive cosmological constant leads to naked bulk singularities akin to those discussed in [30], [31], [33], [34], [35], [36] $-$ but worse, in that the potential singularities would be at a {\it finite} distance from the brane-world.  The wormhole picture replaces the putative singularities by throats sustained by scalar matter which naturally produces the required violation of the bulk SEC; this violation produces the characteristic wormhole ``flare-out'' [24] which averts the formation of the singularities. This suggests that a wormhole in the AdS/CFT bulk is
the correct starting-point.

\ss

Starting with the simplest possible wormhole topology, introducing the requisite ``exotic'' matter, and requiring the absence of naked singularities and various kinds of instability, we have found that two powerful results in global differential geometry (the Witten-Yau-Cai-Galloway theorem and the Gromov-Lawson theorem) have the following gratifying consequences.  First, the wormhole space has to ``fracture'', after the manner of the Randall-Sundrum world; and second, this fracture, our world, must have a positive cosmological constant.  A key role is played here by condition (c) in Theorem 2 above, for, without that condition, the theorem certainly fails.  (The metric (3.2) is smooth everywhere, and it can satisfy conditions (a) and (b), but only at the cost of violating (c).)  Furthermore, it is condition (c) that allows us to rule out a negative (and probably a zero) cosmological constant on the brane.  Yet condition (c) is a purely string-theoretic condition $-$ we need it to control the otherwise unstable process of bulk emission of large branes.  Thus, far from forbidding a positive cosmological constant on the brane-world, string theory apparently {\it requires} it.

\ss
In passing, note that while a bulk with a negative cosmological constant can in principle tolerate a tachyonic scalar field, the same is not true on the brane-world if the latter has a positive cosmological constant.  If, as appears to be the case [52], a tachyonic bulk scalar would induce tachyons on the brane, then we must take care to ensure that the scalar field supporting the wormhole should {\it not} be tachyonic.  Once again, however, this is just the requirement that the scalar potential should be positive; that is, that the bulk SEC should be violated.

\ss

It is clear that the widespread disquiet over the apparent conflict between the observed positivity of the cosmological constant and the predictions of string theory is not to be dispelled by relatively simplistic considerations such as those presented here. These 
constructions do, however, provide a concrete framework for discussing and perhaps resolving
the issues raised in [50] and [51]. For example, our brane-world bulk can be regarded as
having either "finite" volume (for objects which cannot pass through the wormhole throats
to reach the infinite AdS region on the other side) or infinite volume (for those which
can). The problem is to understand precisely how degrees of freedom can ``leak" [53] from
one part of the wormhole space to another.
\ss

If one is inclined to take the wormhole approach more seriously, then the obvious next step is to try to estimate the magnitude (and not just the sign) of the cosmological constant.  As this magnitude is related (see equations (2.15) and (2.16)) to the size of the wormhole throats, and as the latter are supported against collapse by the scalar field, the size of the cosmological constant should be computed from the way the scalar field dynamics keeps the wormhole open.  Another question raised by this work is the following : granted that non-trivial topology in asymptotically AdS spaces has profound physical consequences, what is the CFT dual?  Possibly this is related to the question raised in [54] : what are the bulk degrees of freedom, far from AdS infinity, which are dual to closed Wilson loops in the boundary CFT?

\bs\bs

\nt{\bf REFERENCES}
\ss
\i [1]/ S. M. Carroll, The Cosmological Constant, Living Riviews in Relativity, 
\i /http://www.livingreviews.org/[astro-ph/0004075]

\i [2]/ E. Witten, The Cosmological Constant From the Viewpoint of String Theory.  

\i /[hep-ph/0002297]

\i [3]/ J. Garriga, A. Vilenkin, Solutions to the Cosmological Constant Problems.  

\i /[hep-th/0011262]

\i [4]/ V. Balasubramanian, P. Horava, D. Minic, Deconstructing deSitter.  [hep-th/0103171]

\i [5]/ E. Witten, Quantum Gravity in deSitter Space, talk at Strings 2001, 

\i /http://theory.theory.tifr.res.in/strings/

\i [6]/ A. Chamblin, N. D. Lambert, deSitter space From M-Theory ?  [hep-th/0102159]

\i [7]/ B. McInnes, Topologically Induced Instability in String Theory, JHEP {\bf 0103}(2001) 031 [hep-th/0101136]

\i [8]/ O. Aharony, S. S. Gubser, J. Maldacena, H. Ooguri, Y. Oz, Large N Field Theories, String Theory and Gravity, Phys. Rept. {\bf 323} (2000) 183 [hep-th/9905111]

\i [9]/ L. Randall, R. Sundrum, An Alternative to Compactification, Phys. Rev. Lett. {\bf 83} (1999) 4690.  [hep-th/9906064]

\i [10]/ E. Witten, Comments at ITP Conference on New Dimensions in Field Theory and String Theory, http://online.itp.ucsb.edu/online/

\i [11]/ S. Nojiri, S. D. Odintsov, S. Zerbini, Quantum (In) stability of Dilatonic AdS Backgrounds and Holographic Renormalization Group with Gravity, Phys. Rev. {\bf D62} (2000) 064006.  [hep-th/0001192]

\i [12]/ S. W. Hawking, T. Hertog, H. S. Reall, Brane New World, Phys. Rev. {\bf D62} (2000) 043501.  [hep-th/0003052]

\i [13]/ M. J. Duff, J. T. Liu, Complementarity of the Maldacena and Randall-Sundrum Pictures, Phys. Rev. Lett. {\bf 85} (2000) 2052.  [hep-th/0003237]

\i [14]/ B. McInnes, The Topology of the AdS/CFT/Randall-Sundrum Complementarity, Nucl. Phys. {\bf B602} (2001) 132.  [hep-th/0009087]

\i [15]/ A. L. Besse, Einstein Manifolds, Springer-Verlag, 1987.

\i [16]/ O. De Wolfe, D. Z. Freedman, S. S. Gubser, A. Karch, Modeling the Fifth Dimension with Scalars and Gravity, Phys. Rev. {\bf D62} (2000) 046008.  [hep-th/9909134]

\i [17]/ R. Kallosh, A. Linde, Supersymmetry and the Brane World, JHEP {\bf 0002} (2000) 005.  [hep-th/0001071]

\i [18]/ K. Behrndt, M. Cveti{\v c}, Anti-deSitter Vacua of Gauged Supergravities with 8 Supercharges, Phys. Rev. {\bf D61} (2000) 101901.  [hep-th/0001159]

\i [19]/ A. Ceresole, G. Dall'Agata, Brane-Worlds in 5D Supergravity.  [hep-th/0101214]

\i [20]/ M. J. Duff, J. T. Liu, K. S. Stelle, A Supersymmetric Type IIB Randall-Sundrum Realization  [hep-th/0007120]

\i [21]/ A. Karch, L. Randall, Locally Localized Gravity, Int. J. Mod. Phys. {\bf A16} (2001) 780.  [hep-th/0011156]

\i [22]/ I. I. Kogan, S. Mouslopoulos, A. Papazoglou, A New Bigravity Model with Exclusively Positive Branes, Phys. Lett. {\bf B501} (2001) 140.  [hep-th/0011141]

\i [23]/ G. Gibbons, R. Kallosh, A. Linde, Brane World Sum Rules, JHEP {\bf 0101} (2001) 022.  [hep-th/0011225]

\i [24]/ M. Visser, Lorentzian Wormholes From Einstein to Hawking, AIP Press, 1995.

\i [25]/ E. Witten, Anti-deSitter Space and Holography, Adv. Theor. Math. Phys. {\bf 2} (1998) 253.  [hep-th/9802150]

\i [26]/ B. McInnes, AdS/CFT for Non-Boundary Manifolds, JHEP {\bf 0005} (2000) 025.  [hep-th/000329]

\i [27]/ S. S. Gubser, Curvature Singularities : the Good, the Bad, and the Naked  [hep-th/0002160]

\i [28]/ J. Maldacena, C. Nunez, Supergravity Description of Field Theories on Curved Manifolds and a No-Go Theorem, Int. J. Mod. Phys. {\bf A16} (2001) 822.  [hep-th/0007018]

\i [29]/ S. S. Gubser, AdS/CFT and Gravity, Phys. Rev. {\bf D63} (2001) 084017.  [hep-th/9912001]

\i [30]/ A. Chamblin, G. W. Gibbons, Supergravity on the Brane, Phys. Rev. Lett. {\bf 84} (2000) 1090.  [hep-th/9909130]

\i [31]/ A. Chamblin, S. W. Hawking, H. S. Reall, Brane-World Black Holes, Phys. Rev. {\bf D61} (2000) 065007.  [hep-th/9909205]

\i [32]/ R. Gregory, Black String Instabilities in Anti-de Sitter Space, Class. Quant. Grav. {\bf 17} (2000) L125.  [hep-th/0004101]

\i [33]/ R. Emparan, G. T. Horowitz, R. C. Myers, Exact Description of Black Holes on Branes, JHEP {\bf 0001} (2000) 007.  [hep-th/9911043]

\i [34]/ R. Emparan, G. T. Horowitz, R. C. Myers, Exact Description of Black Holes on Branes II : Comparison with BTZ Black Holes and Black Strings, JHEP {\bf 0001} (2000) 021.  [hep-th/9912135]

\i [35]/ J. Garriga, T. Tanaka, Gravity in the Randall-Sundrum Brane World, Phys. Rev. Lett. {\bf 84} (2000) 2778.  [hep-th/9911055]

\i [36]/ A. Chamblin, C. Csaki, J. Erlich, T. J. Hollowood, Black Diamonds at Brane Junctions, Phys. Rev. {\bf D62} (2000) 044012.  [hep-th/0002076]

\i [37]/ D. Borthwick, Scattering Theory for Conformally Compact Metrics with Variable Curvature at Infinity.  [math.SP/0010273]

\i [38]/ N. Seiberg, E. Witten, The D1/D5 System and Singular CFT, JHEP {\bf 9904} (1999) 017.  [hep-th/9903224] 

\i [39]/ E. Witten, S. T. Yau, Connectedness of the Boundary in the AdS/CFT Correspondence, Adv. Theor. Math. Phys. {\bf 3} (1999) 1635.  [hep-th/9910245]

\i [40]/ A. Ashtekar, A. Magnon, Asymptotically Anti-de Sitter Space-Times, Class. Quant. Grav. {\bf 1} (1984) L39.

\i [41]/ M. Cai, G. J. Galloway, Boundaries of Zero Scalar Curvature in the AdS/CFT Correspondence, Adv. Theor. Math. Phys. {\bf 3} (1999) 1769.  [hep-th/0003046]

\i [42]/ L. Susskind, E. Witten, The Holographic Bound in Anti-deSitter Space.  [hep-th/9805114]

\i [43]/ S. W. Hawking, G. F. R. Ellis, The Large Scale Structure of Space-Time, Cambridge University Press, 1973.

\i [44]/ C. Barcelo, M. Visser, Scalar Fields, Energy Conditions, and Traversable Wormholes, Class. Quant. Grav. {\bf 17} (2000) 3843.  [gr-qc/0003025]

\i [45]/ M. J. Duff, State of the Unification Address.  [hep-th/0012249]

\i [46]/ S. Nojiri, S. D. Odintsov, Supersymmetric New Brane-World.  [hep-th/0102032]

\i [47]/ P. Breitenlohner, D. Z. Freedman, Stability in Gauged Extended Supergravity, Ann. Phys. {\bf 144} (1982) 249.

\i [48]/ H. B. Lawson, M. L. Michelsohn, Spin Geometry, Princeton University Press, 1989.

\i [49]/ S. Nojiri, O. Obregon, S. D. Odintsov, (Non)-singular brane-world cosmology induced by quantum effects in d5 dilatonic gravity, Phys.Rev. {\bf D62} (2000) 104003. [hep-th/0005127]

\i [50]/ W. Fischler, A. Kashani-Poor, R. McNees, S. Paban, The Acceleration of the Universe, a Challenge for String Theory.  [hep-th/0104181]

\i [51]/ S. Hellerman, N. Kaloper, L. Susskind, String Theory and Quintessence.  [hep-th/0104180]

\i [52]/ K. Ghoroku, A. Nakamura, Stability of Randall-Sundrum Brane-World and Tachyonic Scalar.  [hep-th/0103071]

\i [53]/ C. Deffayet, G. Dvali, G. Gabadadze, Accelerated Universe from Gravity Leaking to Extra Dimensions. [astro-ph/0105068]
     
\i [54]/ L. Susskind, N. Toumbas, Wilson Loops as Precursors, Phys. Rev. {\bf D61} (2000) 044001.  [hep-th/9909013]

\bye